\documentclass[12pt]{article}
\nonstopmode
\usepackage[process=auto]{pstool}
\usepackage[x11names, rgb]{xcolor}
\usepackage[utf8]{inputenc}
\usepackage{color}
\usepackage{verbatim}
\usepackage{float}
\usepackage{amssymb}
\usepackage{amsfonts}
\usepackage{amsmath}
\usepackage{graphicx}
\usepackage{mathtools}
\usepackage{draftwatermark}
\SetWatermarkScale{4}
\SetWatermarkLightness{0.90}

\graphicspath{{figures/}{./}}

\usepackage{hyperref}
\hypersetup{
  pdfauthor={Brandon Willard},
  pdftitle={Real-time On and Off Road GPS Tracking},
  pdfsubject={Tracking},
  urlcolor=blue,
}

\DeclareGraphicsExtensions{.pdf,.png,.jpg}
\DeclareUnicodeCharacter{00A0}{~}

\begin{document}

\title{Real-time On and Off Road GPS Tracking}
\author{ Brandon Willard }

\date{
  March 8, 2013\\
  Updated: \today{}
}

\maketitle

\begin{abstract}
This document describes a GPS-based tracking model for position and velocity states
on and off of a road network and it enables parallel, online learning of 
state-dependent parameters, such as GPS error, acceleration error, and road transition
probabilities.  More specifically, the conditionally linear tracking model of \cite{ulmke} 
is adapted to the Particle Learning framework of \cite{lopes}, which provides a foundation 
for further hierarchical Bayesian extensions.  The filter is shown to perform well on a 
real city road network while sufficiently estimating on and off road transition probabilities.
The model in this paper is also backed by an open-source Java project.
\end{abstract}

\section{Introduction}
The availability and ease of collecting GPS data have created a rising interest in
``location aware" applications, many of which define unobservable states such as 
transportation modes (e.g. walking, biking, driving, etc.), traveled roads, and 
a users location/velocity relative to streets, sidewalks or landmarks.  
Real-time GPS data is also being used in trip planning and to ascertain traffic conditions,
which involve similar states and modes.  

Although the GPS accuracy provided by standard devices may be sufficient for
direct use in some simple applications, when one wants to perform inference at
lower sampling
frequencies, or wants to model some underlying discrete state, more is needed.
  For example, when dealing with road networks, there are common cases in which
``snapping" (orthogonally projecting a GPS point onto a street) is not sufficient in itself.  
Often, one needs to at least smooth the resulting sequence of snapped road segments in order 
to get reasonable or coherent results.

Additionally, given the growing use of community sourced maps, such as OpenStreetMap, 
there is an complementary interest in studying the quality of data. 
It is possible to ascertain this value by how well the observed
data obey the restrictions implicit in the street data.  Using a model that
includes on/off-road states, one can identify and inspect high off-road
likelihood data, which can signify problem areas.  

Other use cases for sequential Bayesian estimation involve tracking multiple vehicles with 
varying properties, such that it may be very difficult to batch estimate 
parameter values over historical data, but where it is
possible to define reasonable limits for the parameter values.  
For a system that runs day-to-day and
adds or removes sources of observations regularly (or, similarly, has unreliable
sources), adjustable, sequential parameter learning may be the best option.  Generally, 
when data needs to be removed or augmented real-time, a Bayesian model can easily
provide a coherent framework for such actions. 

 This paper provides an extensible Bayesian model and an efficient particle filter 
for the estimation of its states and parameters through the Particle Learning (PL) framework. 
We establish a conditional dynamic linear model (DLM) for on and off-road 
motion states, and detail the PL estimation procedure.  A method for estimating the 
on-road/off-road probabilities sequentially is provided, and the model is conditionally
amenable to conjugate methods of covariance estimation.

  Many of aforementioned concerns and requirements are well met by techniques within the  
conditional DLM setting of \cite{west}, and PL framework provides sequential approximate
distributions of any estimated parameters.  Commonly, sequential Bayesian applications in 
this domain, have their parameters estimated beforehand, resort to parameter point 
estimates (e.g. Expectation Maximization), or linearizing (e.g. Extended Kalman 
Filters). See \cite{berkely} for an example of the latter, and \cite{liao} for the former. 

In such cases the additional information provided by approximating the full parameter 
distributions isn't available, and/or a departure from a Bayesian setting is made.  Also, 
many such models do not clearly build off of standard motion models, 
which in the case of conditional DLM's are easy to extend in a hierarchical fashion.
For some examples of possible conditional DLM motion models see \cite{motionmodels}.

\section{The Model}
\subsection{Paths and Edges}
An edge is defined as
\begin{align*}
  \lambda &= \left\{(1-\delta)\;l_{\alpha} + \delta\;l_{\omega}\; |\; \delta \in [0,1]\right\}
\end{align*}
for start and end coordinates $l_{\alpha}, l_{\omega} \in \mathbb{R}^2$, and
distance along the edge $\delta \in \mathbb{R}^1$.  
Our construction follows the definition of segments in \cite{ulmke}.

A path is defined as an ordered collection of connected edges 
$\mathcal{P} = \{\lambda^{(N)}, \hdots, \lambda^{(1)}\}$ so that $|\mathcal{P}| =
N$.  It is possible parameterize locations on the path by the distance traveled
along all of the ordered edges, i.e.
\begin{align}
  l(d) &= \begin{cases}
  l_0 + l_1 \frac{d - d_1}{d_1} & 0 \le d \le d_1\\
  l_1 + (l_1 - l_2)\frac{d - d_2}{d_2 - d_1} & d_1 \le d \le d_2\\
  \vdots&\vdots\\
  l_{N-1} + (l_{N-1} - l_N)\frac{d - D}{D - d_{N-1}} & d_{N-1} \le d \le D\\
  \end{cases} \label{eq:segments}
\end{align}
where the distance $d_k$ is given by $d_k = \sum^{k}_{j=1}{|l_{j} - l_{j-1}}|$,
and $l_0=0$.\\

\subsubsection{Evaluated Paths}
  In what follows, we construct our distributions conditional on a path, $\mathcal{P}$.  
Naturally, the results of our model will depend very much on the $\mathcal{P}$ we 
consider and, in application, it is usually impractical to evaluate the entire support of 
$\mathcal{P}$.  Also, it's often clear that the majority of $\mathcal{P}$ will have near 
zero likelihood, so it's reasonable to employ a technique for
finding useful subsets to evaluate, like an efficient sampling procedure or a search
algorithm.  

The approach used here is to employ a modified $A^*$ search between some measure 
of the current state and an area around the observation.  In the DLM-based motion
model one can simply use the observation covariance to determine an area for
finding $A^*$ destination nodes.  As well, other motion model information can
inform the search, such as the projected distance, which can be used as a 
heuristic. 

\subsubsection{Edge Transitions}
The generalized prior distribution for edges is a transition matrix,
where edges, $\lambda^{(l)}$, are interpreted to be a unique integer index and 
$\lambda^{(l)} = 0$ indicates no edge (i.e. off-road).
%\begin{align}
%  p(\lambda^{(j)} | \lambda^{(k)} > 0) &= \begin{cases}
%    \pi_r/N & \lambda^{(j)} \in T(\lambda^{(k)}) > 0\\
%    1-\pi_r & \lambda^{(j)} = 0\\
%    0 & otherwise
%  \end{cases}, \label{eq:edge_trans}\\
%  p(\lambda^{(j)} | \lambda^{(k)} = 0) &= \begin{cases}
%    \pi_f/N & \lambda^{(j)} = 0\\
%    1-\pi_f & \lambda^{(j)} \in T(\lambda^{(k)}) > 0\\
%    0 & otherwise
%  \end{cases} \nonumber
%\end{align}
%where $T(\lambda^{(k)})$ is the set of all edges reachable from 
%$\lambda^{(k)}$, $N=|T(\lambda^{(k)})|$, $\pi_f$, and $\pi_r \in [0, 1]$ are constants.
\begin{align}
  p(\lambda^{(j)} | \lambda^{(k)}, \mathcal{P}) &\propto 
  \begin{cases}
    \pi_{on} & 
      j \geq k \text{ and }
      \lambda^{(j)}  > 0 
      \text{ and } \lambda^{(k)} = 0\\
    \pi_{off} & 
      j \geq k \text{ and }
      \lambda^{(j)} = 0
      \text{ and } \lambda^{(k)} > 0\\
    \pi^{(j,k)}_{r} &  
      j \geq k \text{ and }
      \lambda^{(j)},\lambda^{(k)} > 0\\
    \pi^{(j,k)}_{g} &  
      j \geq k \text{ and }
      \lambda^{(j)} = \lambda^{(k)} = 0\\
    0 & \text{otherwise}
  \end{cases} \label{eq:edge_trans}
\end{align}
for connected edges $\lambda^{(j)}, \lambda^{(k)} \in \mathcal{P}$ and constants 
$\pi_{on}, \pi_{off}, \pi^{(j,k)}_{r}, \pi^{(j,k)}_{g}$.
Basically, we've defined a simple on-off and between-edge transition matrix, for which it is
straight-forward to assign Beta/Dirichlet prior transition probabilities.  

Although we assume a simpler on-off only transition distribution in what follows, resulting in a pair of Beta-Bernoulli, the transition 
matrix formulation makes clear the fact that we can extend this further to cover all road
transitions.

\subsection{Motion Model}
In what follows we specify the probability of assignments to edges
in a path, motion along an edge and free movement.

Given an edge $\lambda > 0$, we track the distance traveled and
velocity along the edges in $\mathcal{P}$.  We define the
\emph{road-movement} state vector as
\begin{align*}
  x^r_i = \begin{pmatrix}
        d_i\\
        v_i\\\end{pmatrix}
\end{align*}
To convert between this vector of road-movement measures and a state vector of 2D
coordinates and 2D velocity, called \emph{ground-movement} coordinates, one
can use the transform
\begin{gather*}
  P_1^\lambda = \frac{l_{\alpha(\lambda)} -
  l_{\omega(\lambda)}}{d_{\omega(\lambda)} - d_{\alpha(\lambda)}},
  \quad
  s_1^\lambda = l_{\alpha(\lambda)} - P^\lambda_1 d_{\omega(\lambda)} \\
  P^\lambda = U (I_2 \otimes P_1^\lambda),
  s^\lambda = U \begin{pmatrix}
  s_1^\lambda\\
  {\bf 0}\\
  \end{pmatrix},
  U = \begin{pmatrix}
    1 & 0 & 0 & 0\\
    0 & 0 & 1 & 0\\
    0 & 1 & 0 & 0\\
    0 & 0 & 0 & 1\\
  \end{pmatrix}
\intertext{where $\alpha(\lambda), \omega(\lambda)$ are indicate the start
and end distances and positions on $\lambda$, and ${\bf 0} = [0, 0]^T$, so that}
  \begin{pmatrix}
    l_{1,i}\\
    v_{1,i}\\
    l_{2,i}\\
    v_{2,i}\\
  \end{pmatrix} = P^\lambda x^r_i + s^\lambda
\end{gather*}
An inverse mapping is \mbox{$d = P^{\lambda,T}(x - s^\lambda)$} for $x \in \mathbb{R}^4$.\\
Since the projection does not preserve the magnitude of velocity when going on-road, 
it is possible that using this transform will have undesired effects in practice.
For instance, projecting onto an edge corresponding to a sharp turn could completely 
diminish the resulting magnitude of velocity.  However, one can simply preserve the 
magnitude of velocity when projecting on-road, or specify another mechanism
for altering the resulting velocity of a projection, if desired.

The transition equations and distribution
\begin{gather}
  x^r_{i+1} = G_r x^r_i + \Gamma_1 \epsilon_{r,i+1}, \; \epsilon_{r,i+1} \sim N(0,
  \sigma^2_r) \label{eq:1d_motion} \\[1em]
  G_r = \begin{pmatrix}
      1 & \Delta t_i  \\
      0 & 1  
    \end{pmatrix}, \;
  \Gamma_r = \begin{pmatrix}
      \frac{{\Delta t_i}^2}{2} \\
      \Delta t_i   
    \end{pmatrix} \nonumber\\
  Var[x^r_{i+1}] \equiv W^r_{i+1} = \Gamma_r \Gamma_r^T \sigma^2_r \nonumber
\end{gather}
For movement not on a path (i.e. $\lambda = 0$), or when referring to ground-movement 
coordinates, we switch to a standard planar position-velocity model with
\begin{gather}
  x^g_{i+1} = G_g x_i + \Gamma_2 \epsilon_{g,i+1}, \; \epsilon_{g,i+1} \sim N(0,
  \sigma^2_{g} I_2) \label{eq:2d_motion} \\[1.0em]
x^g_i = {\begin{pmatrix} l_{1,i}, v_{1,i}, l_{2,i}, v_{2,i} \end{pmatrix}}^T, \; 
  G_g = I_2 \otimes G_r,\;
  \Gamma_g = I_2 \otimes \Gamma_r \nonumber\\[1.0em]
  Var[x^g_{i+1}] \equiv W^g_{i+1} = \Gamma_g \Gamma_g^T \sigma^2_g \nonumber
\end{gather}
All together, we have
\begin{gather*}
  x_i = \begin{cases}
    P^\lambda x^r_i + s^\lambda & \lambda > 0 \\
    x^g_i & \lambda = 0 \\
  \end{cases}
\end{gather*}
To summarize, we have two states, $x^r_i, x^g_i$, straight-line kinematics over
a path and free-movement, respectively, then we constructed a general motion
state $x_i$, which projects the path kinematics to ground-coordinates when
appropriate.\\
Finally, with our generalized motion state, the observation equation is 
\begin{gather*}
  y_i = O_g x_i + \epsilon_{y,i} \\[1em]
  O_g^T = \begin{pmatrix}
      1 & 0   \\
      0 & 0   \\
      0 & 1   \\
      0 & 0   \end{pmatrix},
    \epsilon_{y,i} \sim N(0, \sigma_y^2 I_2) 
\end{gather*}
\subsubsection{Observation and State Covariance Distributions}
A reasonable and effective simplification is to take 
$\sigma_x \equiv \sigma_r = \sigma_g$
and specify the following conjugate priors
\begin{gather*}
  \sigma^2_x \sim IG(\alpha_{x,i}/2, \beta_{x,i}/2)\\
  \sigma^2_y \sim IG(\alpha_{y,i}/2, \beta_{y,i}/2)
\end{gather*}
where $IG(\alpha, \beta)$ specifies an Inverse-Gamma distribution with
shape and scale parameters, respectively.
Like the dynamic factor model example of \cite{lopes}, we can track the
sufficient statistics $\alpha_x, \beta_x, \alpha_y, \beta_y$ through
the conditional conjugacy induced after sampling 
$(x_i, x_{i-1}| y_i, \lambda_i, \sigma_x, \sigma_y)$.
The posterior sufficient statistics are given by
\begin{gather}
  \alpha_{x,i} = \alpha_{x,i-1} + 1\nonumber\\
  \beta_{x,i} = \beta_{x,i-1} 
    + (x_i - G_{\lambda_i}x_{i-1})^T
      (\Gamma_{\lambda_i} \Gamma_{\lambda_i}^T)^{-1} 
      (x_i - G_{\lambda_i}x_{i-1}) \label{eq:covar_posterior_updates}\\
  \alpha_{y,i} = \alpha_{y,i-1} + 1 \nonumber\\
  \beta_{y,i} = \beta_{y,i-1} + (y_i - O_{g}x^g_i)^T(y_i - O_{g}x^g_i) \nonumber
\end{gather}
$G_{\lambda_i}$ signifies whichever state $\lambda_i$ is associated with: $G_r$ or
$G_g$.
%and $p(x_i, x_{i-1} | y_i, \lambda_i, \sigma_x, \sigma_y)$ can be factored into
%$p(x_{i-1} | y_i, \lambda_i) p(x_i | x_{i-1}, y_i, \lambda_i, \sigma_x, \sigma_y)$,
%through straight-forward smoothing steps, and sampled through a Gibbs step.
%The exact distributions are given by
% $(x_{i-1} | y_i, \lambda_i) \sim N(\hat{a}_i, \hat{R}_i),
%  (x_{i}| x_{i-1}, y_i, \lambda_i) \sim N(\hat{h}_i, \hat{H}_i)$
%\begin{gather*}
%  \hat{R}_i = (C_{i-1}^{-1} + G F W^{-1} F^T G^T)^{-1}\\
%  \hat{a}_i = \hat{R}_i (G F W^{-1} y_i + C_{i-1}^{-1} m_{i-1}) \\
%  \hat{H}_i = (F F^T/\sigma_y + (\Gamma \Gamma^T)^{-1}/\sigma_x)^{-1}\\
%  \hat{h}_i = \hat{H}_i (F y_i/\sigma_y 
%    + (\Gamma \Gamma^T)^{-1} x_{i-1}/\sigma_x) 
%\end{gather*}
%for $(x_{i-1} | y_{i-1}, \lambda_{i}) \sim N(m_{i-1}, C_{i-1}),
%W =  \sigma_x^2 F (\Gamma \Gamma^T) F^T$, where $F, G, \Gamma$ are
%implicitly indexed on $\lambda_i$.

\subsubsection{Path Predictions and Posteriors} \label{sec:path_predictions}

We proceed to define the predictive distribution for a path $\mathcal{P}$, and
then the filter updates.  In what follows, path $\mathcal{P}$ always starts with
the previous edge $\lambda_i$.
Also, we define $Z_i$ to be the collection of observations 
and distribution parameters up to observation $i$, 
i.e. $Z_i \equiv \left\{y_i, \mathcal{K}_i, 
\lambda_i, \sigma_x, \sigma_y, 
(\alpha_{x,i}, \beta_{x,i}), (\alpha_{y,i}, \beta_{y,i})  \right\}$ 
where $\mathcal{K}_i$ are
the Kalman filter parameters in ground coordinates  
$\mathcal{K}_i \equiv \left\{m_i, C_i\right\}$ .  

First, let the road-movement prior predictive distribution be
\begin{gather*}
  p\left(x^r_{i+1}| Z_{i} \right) \sim N(\tilde{a}_{i+1}, \tilde{R}_{i+1})
\end{gather*}
which we obtain from \eqref{eq:1d_motion}.  The use of tildes on the mean and
variance parameters signify road-coordinates.\\
A straight-forward construction of the predictive distribution for
a given path would involve strictly assigning the edge that corresponds to
the predicted length, as in
\begin{gather}
  p(x^r_{i+1}, \lambda_{i+1} | \mathcal{P}, Z_i) = p\left(x^r_{i+1}| Z_{i} \right)
      p\left(\lambda_{i+1} | x^r_{i+1}, \mathcal{P} \right)
      p(\lambda_{i+1} | \lambda_{i}, \mathcal{P}) 
    \label{eq:path_prediction} \\
  p(\lambda_{i+1} | x^r_{i+1}, \mathcal{P}) \propto \begin{cases}
    1 & O_r x^r_{i+1} \in [d_{\alpha(\lambda_{i+1})}, d_{\omega(\lambda_{i+1})}] \\
    0 & \text{otherwise} \\
  \end{cases} \label{eq:strict_path}
\end {gather}
where $O_r = \left(1, 0\right)$.  Basically, we want to truncate the state distribution
so that it lies strictly on the given edge; however, such a distribution would forfeit 
the simple conditional linearity and conjugacy, so we resort to a rather effective 
approximation.
Following \cite{ulmke}, we replace the step function in
\eqref{eq:strict_path} by a normal
distribution centered on the edge and with a variance proportional to its
length, 
\begin{gather}
  p\left( \lambda_{i+1} \left| x^r_{i+1}, \mathcal{P} \right.\right) 
\propto f_N\left(\bar{d}_{\lambda_{i+1}} ; O_r x^r_{t+1}, {\Delta d}_{\lambda_{i+1}}^2 \right)
    \label{eq:edge_prob}
\shortintertext{where}
  \bar{d}_\lambda = (d_{\omega(\lambda)} + d_{\alpha(\lambda)})/2, \;
  \Delta d_\lambda = (d_{\omega(\lambda)} - d_{\alpha(\lambda)})/\sqrt{12} \nonumber
\end{gather}
It is important to recall that the distance terms 
$\bar{d}_{\lambda}, \Delta d_{\lambda}, d_{\omega(\lambda)}, d_{\alpha(\lambda)}$ 
are path dependent, such that $d_{\alpha(\lambda)}$ is the distance up to the start
of $\lambda$ on $\mathcal{P}$.  

Moving on, we can derive a categorical prior predictive distribution
over $\lambda_{i+1} \in \mathcal{P}$ using \eqref{eq:path_prediction}
\begin{align*}
  p\left(\lambda_{i+1} | \mathcal{P},  Z_i\right) 
    \propto&\;
  f_N(\bar{d}_{\lambda_{i+1}} ; O_r \tilde{a}_{i+1}, O_r \tilde{R}_{i+1} O^T_r +
{\Delta d}_{\lambda_{i+1}}^2)
\end{align*}

From here we can solve the prior prediction equations, in road-movement
coordinates, for one edge with
\begin{gather*}
  p\left(x^r_{i+1} | \lambda_{i+1}, \mathcal{P}, Z_i\right) 
    \propto p\left(\lambda_{i+1} | x^r_{i+1}, \mathcal{P} \right) 
      p\left(x^r_{i+1} | Z_i\right)
\shortintertext{so that}
  p\left(x^r_{i+1} | \lambda_{i+1}, \mathcal{P}, Z_i\right) \sim 
    N\left(\tilde{a}^\lambda_{i+1}, \tilde{R}^\lambda_{i+1}\right)
\end{gather*}
and, as in \cite{ulmke}%, we use the product formula for normals
\begin{align*}
%  f_N(x;X y, Y) f_N(y;z,Z) = f_N(x;a, A) f_N(y; b, B)\\
%  b = z + W (x - X z), \;a = X z\\
%  B = Z - W A W^T\\
%  W = Z X^T A^{-1}\\
%  A = X Z X^T + Y\\
%\shortintertext{and obtain}
  \tilde{a}^\lambda_{i+1} &= \tilde{a}_i 
  + W^\lambda_{i+1}\left(\bar{d}_\lambda - O_r \tilde{a}_i \right), &
  \tilde{R}^\lambda_{i+1} &= \tilde{R}_{i+1} 
    - W^\lambda_{i+1} S^\lambda_{i+1} {W^\lambda_{i+1}}^T\\
    W^\lambda_{i+1} &= \tilde{R}_{i+1} O_r^T S^{-1,\lambda}_{i+1}, & 
    S^\lambda_{i+1} &= O_r \tilde{R}_{i+1} O_r^T + {\Delta d_{\lambda}}^2
\end{align*}
We can now obtain the prior predictive distribution in ground-coordinates
\begin{gather*}
  p\left(x_{i+1} | \mathcal{P}, Z_i \right) \propto 
    \sum_{\lambda_{i+1} \in \mathcal{P}}
    { p\left(x_{i+1} | \lambda_{i+1}, \mathcal{P}, Z_i\right)
      p\left(\lambda_{i+1} | \mathcal{P}, Z_i \right)
      p(\lambda_{i+1} | \lambda_{i}, \mathcal{P}) 
      }\\
  p\left(x_{i+1} | \lambda_{i+1}, \mathcal{P}, Z_i \right) \sim 
    N\left(a^\lambda_{i+1} =P^\lambda \tilde{a}^\lambda_{i+1} + s^\lambda, 
      R^\lambda_{i+1} 
        = P^\lambda \tilde{R}^\lambda_{i+1} P^{\lambda,T} \right)
\end{gather*}
\\
Filtering is accomplished by conversion to ground coordinates, then updating the
state sufficient statistics, through normal Kalman filtering
\begin{align*}
  p\left( x_{i+1} |\mathcal{P}, Z_{i+1} \right) &\propto 
    \sum_{\lambda_{i+1} \in \mathcal{P}} {
      p\left(\lambda_{i+1} | \lambda_{i}, \mathcal{P}, Z_{i} \right) 
      \int_{x_{i+1}}{
        p\left(y_{i+1} | x_{i+1} \right) 
        p\left(x_{i+1} | \lambda_{i+1}, \mathcal{P}, Z_{i} \right)
      }
    }\\
  &\propto \sum_{\lambda_{i+1} \in \mathcal{P}} {
    p\left(\lambda_{i+1} | \lambda_{i}, \mathcal{P}, Z_{i} \right) 
    p\left(x_{i+1} | \lambda_{i+1}, \mathcal{P}, Z_{i+1} \right) 
  }                                                             
\end{align*}
where
\begin{align}
  p\left(\lambda_{i+1} | \lambda_i, \mathcal{P}, Z_{i} \right)
    &= p\left(\lambda_{i+1} | \mathcal{P}, Z_{i} \right) 
      p(\lambda_{i+1} | \lambda_{i}, \mathcal{P})\nonumber\\
      p\left(x_{i+1} | \lambda, Z_{i+1} \right) &\sim
    N\left(m^\lambda_{i+1}, C^\lambda_{i+1}\right) \nonumber
\shortintertext{with posterior mean and covariance}
  m^\lambda_{i+1} =& 
    a^\lambda_{i+1} + A^\lambda_{i+1} (y_{i+1} - e^\lambda_{i+1})
    \label{eq:posterior_updates}\\
  C^{-1,\lambda}_{i+1} =& R^{-1,\lambda}_{i+1} + O_g^T O_g/\sigma^2_y \nonumber\\
  A^\lambda_{i+1} =& R^\lambda_{i+1} O^T_g Q^{-1,\lambda}_{i+1}\nonumber
\shortintertext{and predictive mean and covariance}
  e^\lambda_{i+1} =& O_g a^\lambda_{i+1}\nonumber\\
  Q^\lambda_{i+1} =& O_g R^\lambda_{i+1} O^T_g + \sigma^2_y I_2\nonumber
\end{align}
and then conversion back to edge coordinates
\begin{align*}
  p\left( x^r_{i+1} | \mathcal{P}, Z_{i+1} \right) 
    \propto \sum_{\lambda_{i+1} \in
      \mathcal{P}}
      & p\left(\lambda_{i+1} | \lambda_{i}, \mathcal{P}, Z_{i+1} \right) \\
      & \times N\left(\tilde{m}^\lambda_{i+1} = P^{\lambda,T} (m^\lambda_{i+1} - s^\lambda), 
      \tilde{C}^\lambda_{i+1} = P^{\lambda,T} C^\lambda_{i+1} {P^\lambda} \right)
\end{align*}
Finally, for off-road movement, filtering is performed in ground 
coordinates, using the standard Kalman filter updates.

\section{Estimation}

Since our state, $x_i$, is a DLM conditional on a path and edge, we are able to 
use Rao-Blackwellization and sufficient statistics.  The likelihoods and
posteriors are conditionally known, and we are able to 
sample directly from them, so the results are perfectly adapted (\cite{pitt}, \cite{lopes}).
This provides a setting that is well suited for estimating hyper-parameters online, even
when their estimation entails sampling and not simply conjugate updates.  

Considering the defined distributions, we have a situation very 
reminiscent of the tracking example in \cite{Liu_SISR}, and the dynamic factor
model with time-varying loadings in \cite{lopes}, which includes estimation for
observation and state covariances.  
Through a similar design we attempt to relate their studied efficiencies and improvements 
to components of our model.
Specifically, \cite{Liu_SISR} demonstrates improvements of mixture Kalman filter (MKF)
state estimation over plain SIS or particle filters, and \cite{lopes} shows the efficiency 
of state-transition and variance estimation in the PL setting for a MKF.
These results have motivated our use of MKF/PL in what follows.\\

We follow the CDLM section and switching model example in \cite{lopes}. 
Starting with $N$ particles of 
$Z_i = \left\{x_i, \lambda_i, \mathcal{K}_i, \sigma_x, \sigma_y, 
(\alpha_{x,i}, \beta_{x,i}), (\alpha_{y,i}, \beta_{y,i}) \right\}$
approximating $p\left(x_i, \lambda_i, \mathcal{K}_i| y_i \right)$ the particle
filter steps are
\begin{enumerate}
  \item \emph{Re-sampling:} Draw an index $k^j \sim Multi(w_i^{(1)}, \dots, w_i^{(N)})$
  with weights 
  \begin{gather*}
    w_i^{(j)} \propto p(y_{i+1} | (\mathcal{K}_i, \lambda_i)^{(k^j)})
  \shortintertext{with}
  p(y_{i+1} | \mathcal{K}_i, \lambda_i) = 
      \sum_{\mathcal{P}_{i+1}}{
        \sum_{\lambda_{i+1} \in \mathcal{P}_{i+1}}{
          f_N(y_{i+1} ; e^{\lambda_{i+1}}_{i+1}, Q^{\lambda_{i+1}}_{i+1}) 
          \;p(\lambda_{i+1} | \lambda_i, \mathcal{P}_{i+1}, Z_{i})
          }
        }
  \end{gather*}

  \item \emph{Propagating state $\lambda$:}  Draw $\lambda_{i+1}^{(j)}$ 
  from
  \begin{gather*}
    p(\lambda_{i+1} | (\mathcal{K}_i, \lambda_i, \mathcal{P}_{i+1})^{(k^j)}, y_{i+1}) 
    \propto f_N(y_{i+1} ; e^{\lambda_{i+1}}_{i+1}, Q^{\lambda_{i+1}}_{i+1}) 
          \;p(\lambda_{i+1} | \lambda_i, \mathcal{P}_{i+1}, Z_{i})
  \end{gather*}

  \item \emph{Propagating state $x_{i+1}$:}  Draw $x_{i+1}^{(j)}$ from 
    $p(x^{(j)}_{i+1} | \lambda_{i+1}, y_{i+1}, \mathcal{K}_i, \sigma_x, \sigma_y)$

  \item \emph{Propagating state sufficient statistics, $\mathcal{K}_{i+1}$:} Perform the
  Kalman filter recursions described in \eqref{eq:posterior_updates}, or, for 
  $\lambda_{i+1} = 0$, the standard Kalman recursions.

  \item \emph{Propagating parameter sufficient statistics:} \label{list:suff_stats} 
    Perform off-line updates for hyper-parameters
    $\pi_{on}, \pi_{off}$ and \eqref{eq:covar_posterior_updates}, 
    then draw $\sigma_x, \sigma_y$.
\end{enumerate}

\section{Implementation}
In what follows we observe the performance of the Particle Learning filter
described above against a simple Bootstrap filter.  We are primarily concerned
with basic accuracy, stability and the computational requirements for 
runs of decent length over a non-trivial graph.  
Secondly, through simulation we are able to directly measure the accuracy of
parameter estimation.
For our implementation we used a graph of Washington DC, and generated
roughly 1000 of observations for each run.

  The Bootstrap filter estimates only the states, as described below, and the 
PL filter estimates the states and edge transitions.
To do this, the distribution in \eqref{eq:edge_trans} is partitioned into two Bernoulli
distributions with Beta distributed transition probabilities, under the restriction that 
$\pi_{r}^{(j,k)} = \pi_{r}, \pi_{g}^{(j,k)} = \pi_{g}$ (the same restriction is
used in the simulations and bootstrap filter).  Due to conjugacy, updates are
straight-forward and performed in step \ref{list:suff_stats} above.  This construction
follows the transition matrix in Example 3 of \cite{lopes}.

\subsection{Bootstrap Filter and Simulation Procedure}
The bootstrap filter (BS) consists of sampling $p(x_{i+1} | x_i, \mathcal{P})$ by first
updating per \eqref{eq:1d_motion} or
\eqref{eq:2d_motion}, and including the error due to the state transition
covariance.  Then, for the case of $\lambda \ne \emptyset$ we move in
the direction and length of the sampled movement, sampling edge 
transitions according to \eqref{eq:edge_trans} and \eqref{eq:strict_path}. 
Simulations are produced in exactly the same way.

Furthermore, when filtering, the effective sample size (ESS) is 
calculated for the propagated sample set and a condition is checked to 
determine if resampling is required.  The condition was set to $ESS < 0.9 N$,
where $N$ is the initial sample size.

\subsection{Setup}
Both models are run on the same simulated data set with varying particle sizes
and run lengths.  The true parameters were
\begin{gather*}
  \Delta t = 30\\
  \sigma^2_y = 100\, I_2\\
  \sigma^2_r = 6.25 \times 10^{-4}\\
  \sigma^2_g = 6.25 \times 10^{-4} I_2\\
  \pi_g = 0.05, \pi_{on} = 0.95\\
  \pi_{off} = 0.05, \pi_r = 0.95
\end{gather*}  
The BS filter was run using the true parameters as well as the PL filter with
the exception of the $\pi$ parameters, which were replaced by Beta parameters equaling    
\begin{gather*}
  (\alpha_{off,off} = 15, \alpha_{off,on} = 20) \\
  (\alpha_{on,on} = 70, \alpha_{on,off} = 100)
\end{gather*}  
for the ``off" and ``on" distributions, respectively.

Simulations were performed on an Intel Core i5 
with 16 gigs of RAM on Ubuntu 12.01 and using Java 6 OpenJDK amd64, and for each 
observation the root mean squared error (RMSE) of the position-velocity
state vector, in ground-coordinates, was computed over the set of posterior particles.

\subsection{Results}
\begin{figure}[htb]
  \centering
  \includegraphics[width=\linewidth]{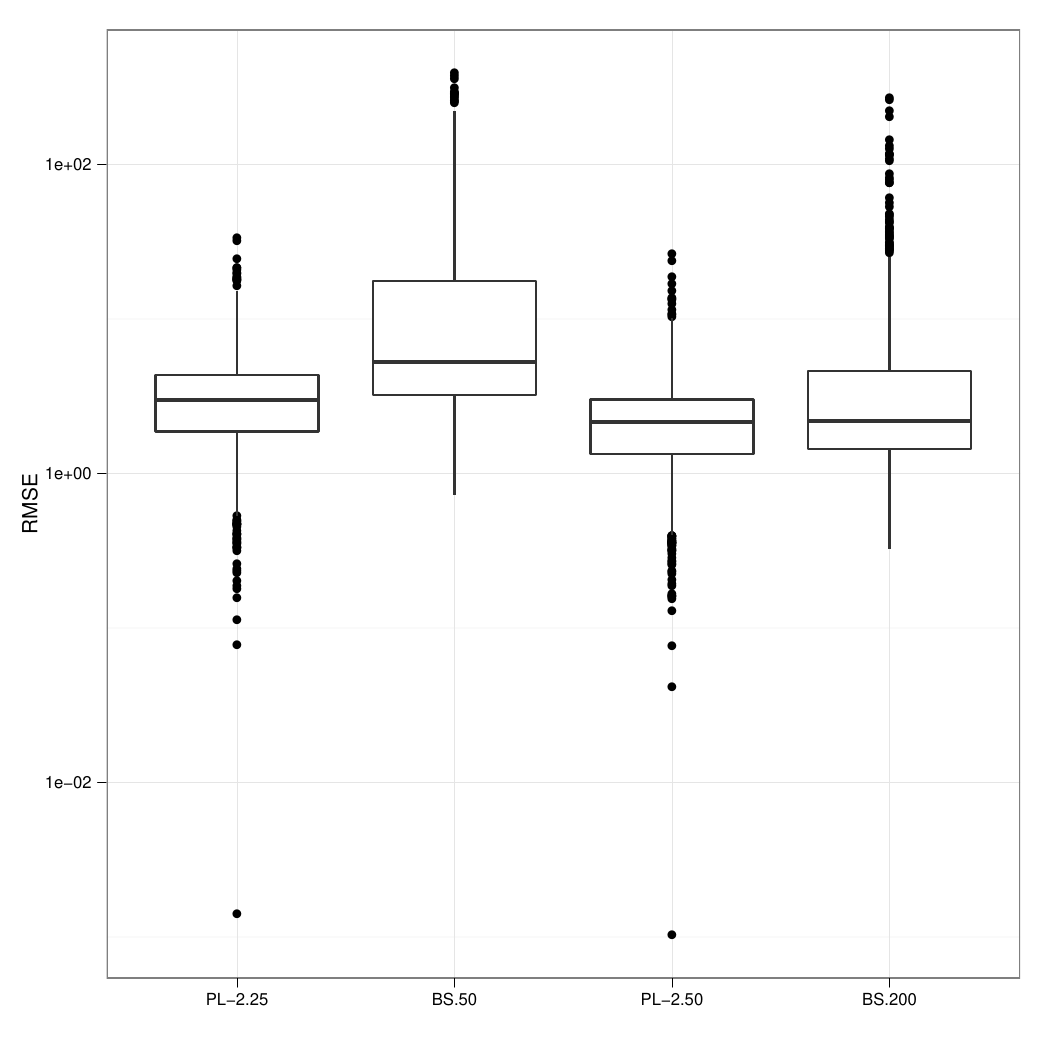}
  \caption{RMSE Comparisons}
  \label{fig:plbs-rmse-1200-wisker}
\end{figure}

\begin{figure}[htb]
  \centering
  \includegraphics[width=\linewidth]{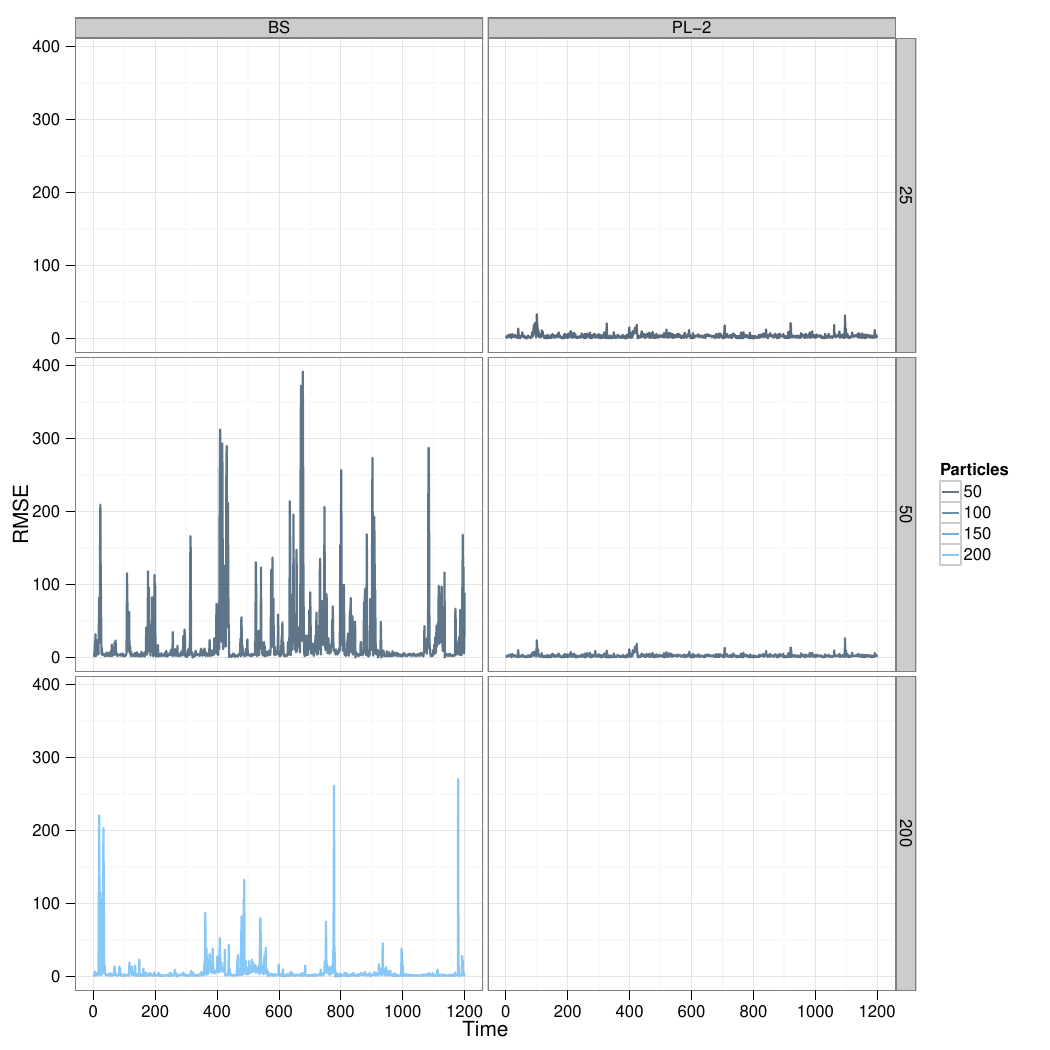}
  \caption{RMSE time series.}
  \label{fig:plbs-rmse-1200-series}
\end{figure}

\begin{figure}[htb]
  \centering
  \includegraphics[width=\linewidth]{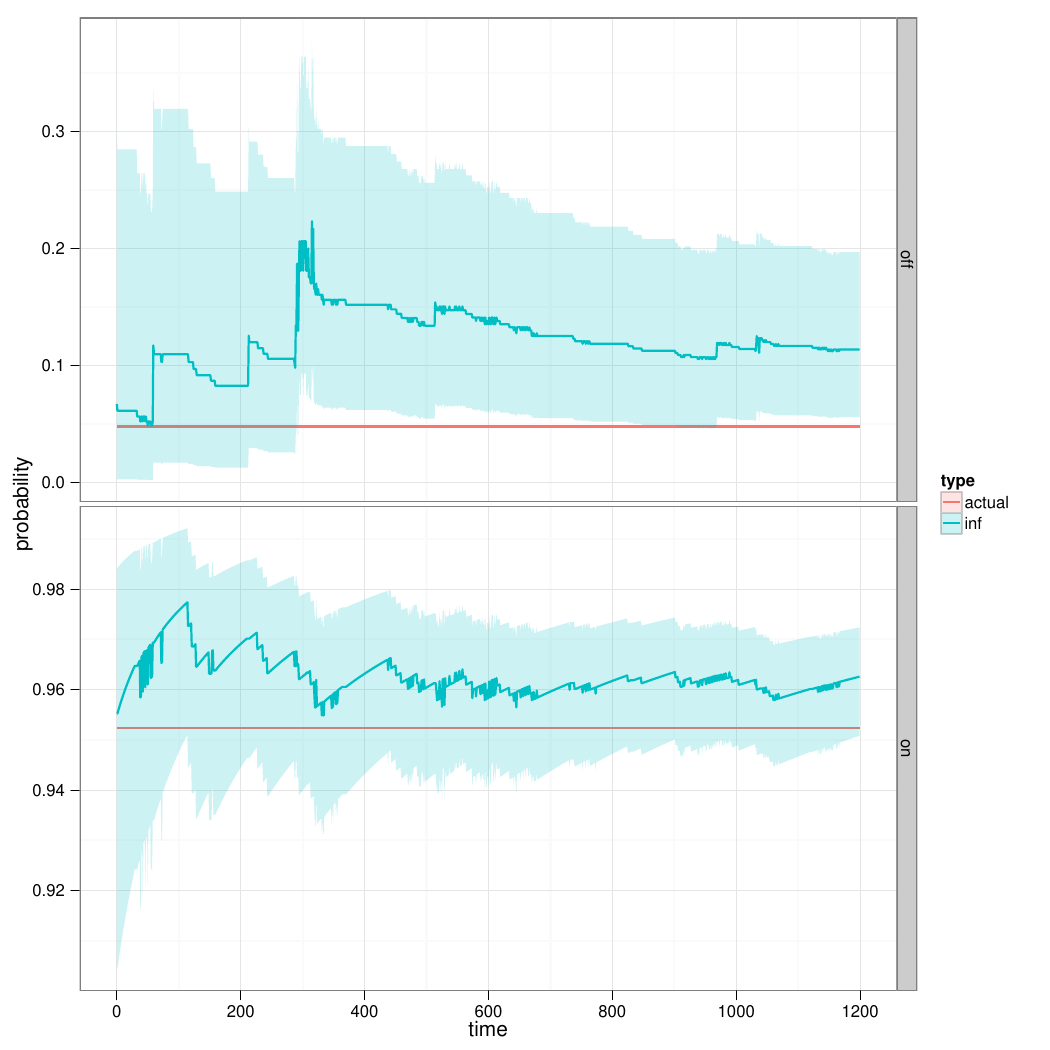}
  \caption{Posterior Median 95\% Credibility Intervals for $\pi_{off}, \pi_{on}$}
  \label{fig:pl-edge-trans-1200-95CI}
\end{figure}

\begin{figure}[htb]
  \centering
  \includegraphics[width=\linewidth]{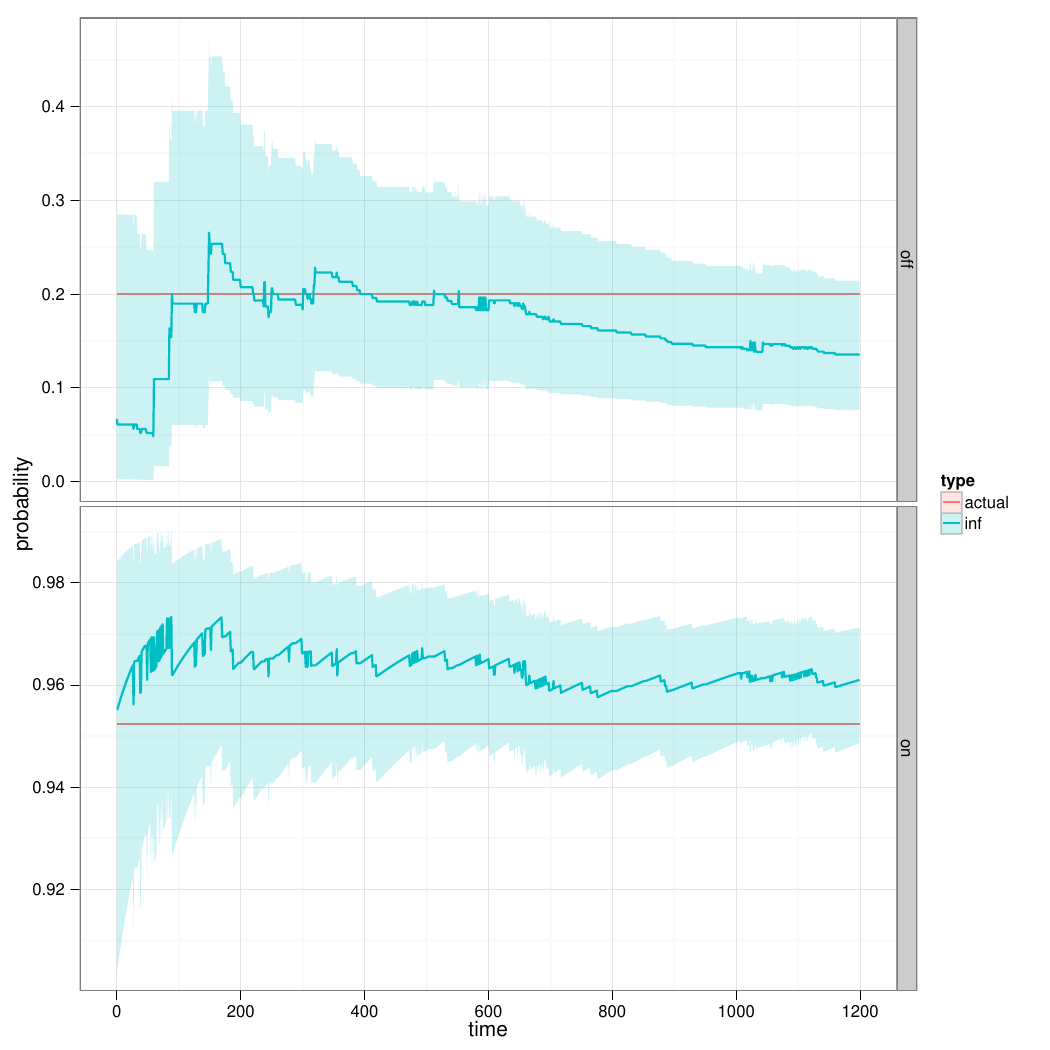}
  \caption{Posterior Median 95\% Credibility Intervals for $\pi_{off}, \pi_{on}$}
  \label{fig:pl-edge-trans-1200-95CI-2}
\end{figure}

Figures~\ref{fig:plbs-rmse-1200-wisker},~\ref{fig:plbs-rmse-1200-series} compare the log
RMSE values for the two filters.  Figure
~\ref{fig:plbs-rmse-1200-wisker} shows that PL RMSE is lower, on average, than
BS, across all particle sizes, but the most noticeable difference is how many
fewer particles PL takes for its accuracy.  Figures~\ref{fig:pl-edge-trans-1200-95CI-2},
~\ref{fig:pl-edge-trans-1200-95CI} show 95\% credibility intervals for the estimated
$\pi_{off,off}, \pi_{on,on}$ with $\sigma^2_g = 6.25 \times 10^{-3} I_4$ in both the 
simulations and PL filters.

  The online estimation of $\pi_{off,off}, \pi_{on,on}$ demonstrates some of the inherent
complexities with having on and off-road states.  
For example, in Figure~\ref{fig:pl-edge-trans-1200-95CI} we can see that 
estimation is biased upward in both terms.  Observing the data directly, we often find
that off-road states are not propagated due to the location in which the simulation
goes off-road.  Where there are many streets close together, an off-road state
that lasts for one observation is
easy to interpret as an on-road state, since there will likely exist connecting
roads that fit with the true velocity and location.  The same goes for cases in which an 
off-road state is traveling parallel to or, for all practical purposes, on a road.
We would expect that longer off-road sojourn times would help differentiate, e.g.
Figure~\ref{fig:pl-edge-trans-1200-95CI}, or noticeably different dynamics for
off-road states, and they do. The off-road to off-road probability in
Figure~\ref{fig:pl-edge-trans-1200-95CI} is very small, so we can expect
that staying off-road is very brief, and, given the large size of $\pi_{on}$,
the simulation doesn't go off-road very often.  Figure~\ref{fig:pl-edge-trans-1200-95CI-2}  
shows that increasing $\pi_{off}$ can help identify the signal.

  %Table \ref{table:plbs-times} displays the throughput for each filter.  
The results were computed in Java, synchronously, and, 
are naturally implementation dependent.  
Most of the time spent in the PL filter 
revolves around an $A^*$ computation of
shortest paths.  In the implementation used here, some caching was utilized to
avoid repeated computations, and, of course, the BS filter does not require such
computations.
The BS filter is clearly the faster choice for low particle size, but referencing Figures
\ref{fig:plbs-rmse-1200-wisker}, \ref{fig:plbs-rmse-1200-series} we can see that
to achieve an RMSE comparable to the PL filter one needs to increase the particle
size, which pulls the BS filter's throughput results to those of the PL filter at 25
particles.

  Regarding the filter comparison in generality, we're mostly seeing the
documented benefit of using fully-adapted, Rao-Blackwellized filters, since
the advantage of path-searching over path-sampling has been mostly diminished.
To address this difference, the above
tests were performed with parameters that restricted the distances moved between 
observations
to values that do not push the limits of path-searching against sampling.
The simulations spend most of their time transitioning between 0-3 edges, so
that simple edge sampling is unlikely to fail at finding the true paths.  

  For models or data with larger
state covariances, and thus more error due to acceleration, the comparison between
path generating methods may not be useful, as path-sampling models like the BS filter
will quickly face issues.  The same goes for settings in 
which the graph
is more dense and/or paths are more restricted. In tests, when dead-end's are reached, or
highly divergent paths are taken, we've observed that both models will eventually go off-road
and find their way back to the correct path (this also accounts for some of the 
bias in $\pi_{on},\pi_{off}$ estimation).  This can be seen in the RMSE results, 
as in \ref{fig:plbs-rmse-1200-series}, where large values appear in series.  Even though
the simulation parameters are less likely to cause these events, the city graphs are 
non-trivial, so the behaviour is present, and, for the times in which it does occur, we 
can see that with increasing particle size the BS filter is less affected, as expected.   

\section{Conclusion and Future Work}

  The results demonstrate that the PL model is effective in
this setting, and that it can perform well relative to a standard bootstrap filter based
on the exact data generating process for an actual city graph.
The simulations were designed to approximate the observed on and off road
behaviour of real data collected in Cebu and DC, and the estimation was 
sufficient for those settings.  
Generally, outside of the present paper's simulations and in real use cases,
the PL filter's MAP results were observed to be overall accurate
at inferring traveled segments and on/off states containing a mix of walking 
and biking in the DC area.  In summary, alongside the know benefits of MKF filtering in
a tracking scenario, we can add the ability to infer traveled paths and off-road behaviour
in real-time.

  Future work could involve demonstrating and studying more of the
learning aspects, like covariance estimation, by using more interesting edge
transition models, as well as testing the limits of accuracy in
varied acceleration settings.  Comparisons with other sophisticated or
specialized models may be of use in determining comparative advantages
in certain scenarios, too.  Otherwise, one clear line of investigation involves
an analysis of the resulting smoothed paths, since many applications of such a
model require accurate edge-to-edge estimation, or even the elucidation of
off-road trajectories.

  The model is available as an open-source project at 
\url{https://github.com/brandonwillard/open-tracking-tools}.  The results in this paper
were all produced with the aforementioned code, which is setup to filter custom
csv data and produce simulations for any city using OpenTripPlanner graphs
(\url{opentripplanner.org}) built with OpenStreetMap
(\url{www.openstreetmap.org}) data.  

\section{Thanks}
I would like to thank Prof. Yali Amit, Mei Wang and Hedibert Lopes for their input and 
support, and OpenPlans.org for the great community and opportunities to apply this work.

%\begin{table}
%  \begin{center}
%    \begin{tabular}{ | l | c |}
%      \hline 
%      Filter & Records/Second \\ \hline
%       PL 25 & 67.35\\
%       PL 50 & 39.14\\
%       PL 200 & 11.59\\
%      \hline
%       BS 50 & 1673.64\\
%       BS 200 & 215.74\\
%       BS 500 & 51.49\\
%      \hline
%    \end{tabular}
%    \caption{Filter Throughput}
%    \label{table:plbs-times}
%  \end{center}
%\end{table}

%
%
%  Figures 
%
%

\end{document}